\documentclass[twocolumn,prl,aps,superscriptaddress,10pt]{revtex4-2}
\usepackage[T1]{fontenc}
\usepackage[latin9]{inputenc}
\usepackage{amsmath,amsfonts}
\usepackage{float}
\usepackage{graphicx}
\usepackage[english]{babel}
\usepackage[bookmarks=true, colorlinks, linkcolor=blue, urlcolor=blue, citecolor=blue, plainpages=false, pdfpagelabels, final, breaklinks=true]{hyperref}

\begin{document}

%% Some macros
\newcommand\bra[1]{\mathinner{\langle{\textstyle#1}\rvert}}
\newcommand\ket[1]{\mathinner{\lvert{\textstyle#1}\rangle}}
\newcommand\braket[1]{\mathinner{\langle{\textstyle#1}\rangle}}
\newcommand\hata{\hat{a}}
\newcommand\hatb{\hat{b}}
\newcommand\hatc{\hat{c}}
\newcommand\hatH{\hat{H}}
\newcommand\hatN{\hat{N}}
\newcommand\hatU{\hat{U}}
\newcommand\hatV{\hat{V}}
\newcommand\hatW{\hat{W}}
\newcommand\calC{\mathcal{C}}
\newcommand\calD{\mathcal{D}}
\newcommand\calH{\mathcal{H}}
\newcommand\calS{\mathcal{S}}
\newcommand\calZ{\mathcal{Z}}
\newcommand\msc[1]{\textbf{\textcolor{red}{MSC: #1}}}

\title{Geometric Manipulation of a Decoherence-Free Subspace in 
Atomic Ensembles}

\author{Dongni Chen}
\affiliation{Institute of Theoretical Physics, Chinese Academy of Sciences, Beijing 100190, People's Republic of China}
\affiliation{School of Physical Sciences, University of Chinese Academy of Sciences, Beijing 100049, People's Republic of China}

\author{Si Luo}
\affiliation{Beijing Computational Science Research Center, Beijing 100193, People's Republic of China}

\author{Ying-Dan Wang}
\affiliation{Institute of Theoretical Physics, Chinese Academy of Sciences, Beijing 100190, People's Republic of China}
\affiliation{School of Physical Sciences, University of Chinese Academy of Sciences, Beijing 100049, People's Republic of China}

\author{Stefano Chesi}
\email{stefano.chesi@csrc.ac.cn}
\affiliation{Beijing Computational Science Research Center, Beijing 100193, People's Republic of China}
\affiliation{Department of Physics, Beijing Normal University, Beijing 100875, People's Republic of China}

\author{Mahn-Soo Choi}
\email{choims@korea.ac.kr}
\affiliation{Department of Physics, Korea University, Seoul 02841, South Korea}

\begin{abstract}
We consider an ensemble of atoms with $\Lambda$-type level structure trapped
in a single-mode cavity, and propose a geometric scheme of coherent
manipulation of quantum states on the subspace of zero-energy states within
the quantum Zeno subspace of the system. We find that the particular subspace
inherits the decoherence-free nature of the quantum Zeno subspace and features
a symmetry-protected degeneracy, fulfilling all the conditions for a universal
scheme of arbitrary unitary operations on it.
\end{abstract}

\maketitle

%%%%%%%%%%%%%%%%%%%%%%%%%%%%%%%%%%%%%%%%%%%%%%%%%%%%%%%%%%%%%%%%%%%%%%%%%%%%%%
% \section{Introduction}

Coherent manipulation of quantum states is an essential part of various
quantum technologies ranging from quantum-enhanced precision measurement to more
ambitious goals, like quantum simulation and quantum information processing.
% Unfortunately, decoherence effects and operational imperfections have
% imposed huge challenges on it.
While there remain considerable challenges to achieve reliable quantum-state
engineering on large scales \cite{Preskill18a}, a number of schemes to suppress
and/or control decoherence and improve operational imperfections have been
proposed and are currently under investigation. Notable examples include
approaches based on decoherence-free subspaces \cite{Beige2000_NJP2-22,
  Beige2000_PRL85-1762, Lidar1998_PRL81-2594}, dynamical decoupling
\cite{Viola1998_PRA58-2733, Vitali1999_PRA59-4178, Zanardi1999_PLA258-77},
quantum error correction \cite{Shor1995_PRA52-R2493, Steane1996_PRL77-793,
  Plenio1997_PRA55-67}, and holonomic manipulation
\cite{Zanardi1999_PLA264-94, Jones2000_Nature403-869,
  Duan2001_Science292-1695, Sjoqvist12a, Toyoda2013_PRA87-052307,
  Feng2013_PRL110-190501, Abdumalikov2013_Nature496-482, Zu14a,
  Xu2018_PRL121-110501, Duan2019_PRL122-010503}.
In addition, topological approaches \cite{Freedman1998_PNAS95-98,
  Kitaev2003_AnnalsofPhysics303-2,Nayak08a} have recently attracted remarkable
interest, due to the highly appealing prospect of topologically
protected operations.  However, physical systems with robust and easily addressable
topological entities are yet to be discovered or developed
\cite{Mourik12a,Deng12a,Das12a,Nadj-Perge14a,Bartolomei20a}.

In this work, we combine the self-correcting features of geometric methods and
the concept of decoherence-free subspaces. More specifically, we develop a
holonomic manipulation scheme based on ensembles of atoms with $\Lambda$-type
level structure, trapped in a single-mode cavity \cite{endnote:1}. To
implement universal holonomies, a crucial requirement is the degeneracy of the
operational subspace (or, the equally-demanding cyclic-evolution condition
\cite{Aharonov87a, Sjoqvist12a}). Here we identify a generally degenerate
subspace, whose decoherence-free feature is inherited from the quantum Zeno
subspace \cite{Facchi2002_PRL89-080401}. The atomic systems considered here
have been popular for studies of dissipation-based quantum computation
\cite{Beige2000_PRL85-1762}, quantum gates based on adiabatic passages (or
shortcuts to them) \cite{Wu2019_JPA52-335301,Song2016_NJP18-023001},
non-Adiabatic holonomic quantum computation \cite{Mousolou2018_JPA51-475303},
and generation of highly entangled states \cite{Yang2010_NJP12-113039,
  Shao2010_EPL90-50003, Chen2014_QIP13-1857, Wu2017_SciRep7-46255}. However,
the degeneracy allowing for holonomic manipulation, not to speak of the
combination with the decoherence-free character of the quantum Zeno subspace,
has not been exploited in those works. By focusing on the simplest realization
of our scheme, we prove the universality of the holonomic gates. Combining the
self-correcting character of unitary operations implemented through geometric
methods with the decoherence-free feature of this particular subspace might
effectively provide a high level of fault tolerance, practically comparable to
topological methods.

% \section{Degenerate Decoherence-Free Subspace}
\paragraph{Degenerate Decoherence-Free Subspace.}
\label{Sec:Model}

\begin{figure}
\centering
\includegraphics[width=80mm]{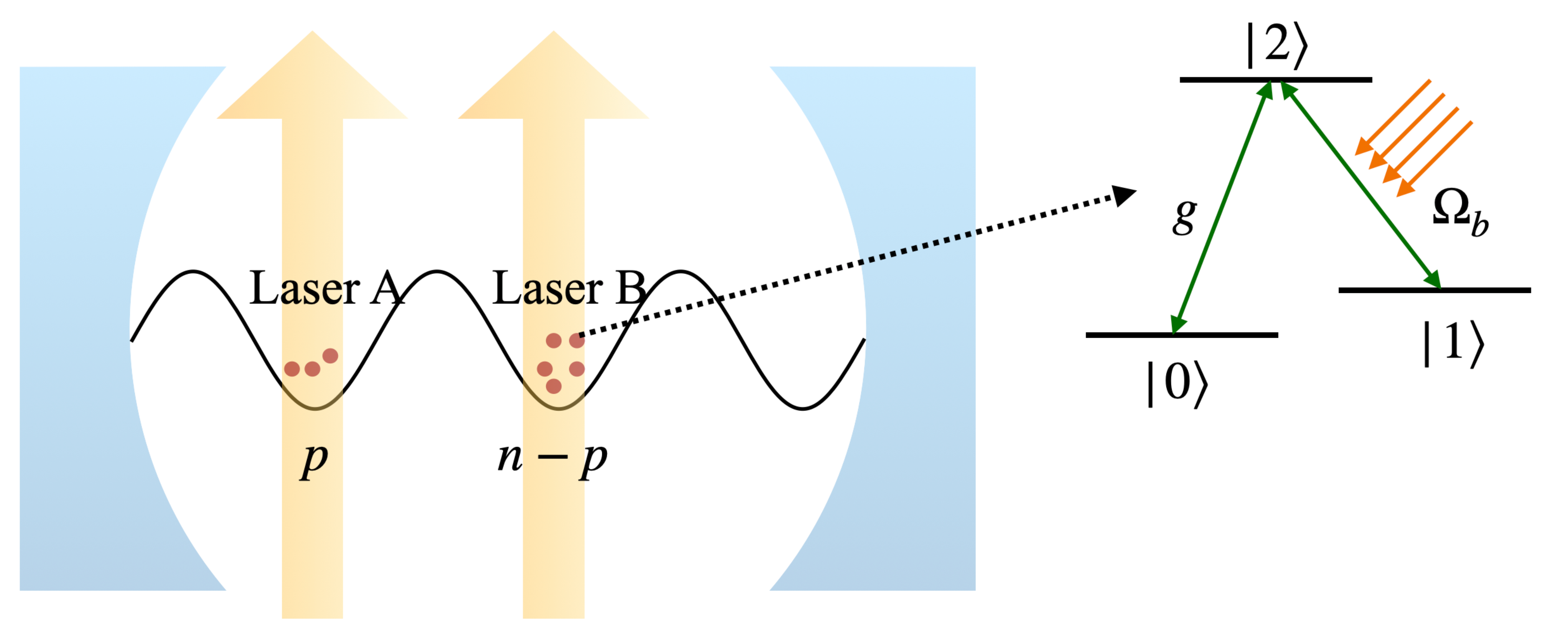}       
\caption{Schematics of the atomic ensembles in a single-mode cavity (left),
and the level structure of each atom (right).}
\label{model}
\end{figure}

We consider $n$ identical atoms with $\Lambda$-type level structure
\cite{endnote:1} inside a single-mode cavity (see Fig.~\ref{model}).
The two ground states of each atom are denoted by $\ket{0}$ and $\ket{1}$,
respectively, and the excited state by $\ket{2}$.  The transition
$\ket{0}\leftrightarrow\ket{2}$ is induced by the cavity photon while the
transition $\ket{1}\leftrightarrow\ket{2}$ is driven resonantly by an external
classical field.  All atoms are assumed to be coupled to the cavity photon
with a uniform strength $g$.  We divide the atoms into two subensembles, $A$
containing $p$ atoms and $B$ with the rest, and separately tune their
characteristic Rabi transition amplitudes, $\Omega_a$ and $\Omega_b$,
respectively (atoms in each subensemble have a uniform Rabi transition
amplitude).  In the interaction picture, the dynamics of the system is
governed by the Hamiltonian $\hatH=\hatH_g+\hatH_\Omega$ with
\begin{subequations}
\label{Hamiltonian:1}
\begin{align}
\label{Hamiltonian:1a}
\hat H_{g}= & g\sum_{j=1}^{n}\hat{c}|2\rangle_{j}\langle0|+\mathrm{H.c.}\,,\\
\label{Hamiltonian:1b}
\hat H_{\Omega}= & \Omega_{a}\sum_{j=1}^{p}|2\rangle_{j}\langle1|
+ \Omega_{b}\sum_{j=p+1}^{n}|2\rangle_{j}\langle1|+\mathrm{H.c.} \,,
\end{align}
\end{subequations}
where $\hat{c}$ is the annihilation operator of the cavity photon and
$\ket{s}_j$ denotes the $j$th atom in state $\ket{s}$ ($s=0,1,2$).
 
To identify the relevant decoherence-free subspace,
% Before going further,
we exploit the symmetries in the Hamiltonian~\eqref{Hamiltonian:1}:
% and simplify the arguments later on:
First, the total excitation number
\begin{math}
\hat{N} = \hat c^\dag \hat c +
\sum_j(|1\rangle_j\langle 1| + |2\rangle_j\langle 2|)
\end{math}
is conserved, $[\hat{N},\hat{H}]=0$.  It allows us to focus on an invariant
subspace $\mathcal{H}_{N}$ with a particular excitation number $N$.
Throughout this work, the excitation number is assumed equal to the number $p$
of atoms in subensemble $A$, which simplifies the initial preparation of the
system.
Second, the Hamiltonian is invariant under exchange of any pair of atoms
within each subensemble.  Among various invariant subspaces, we are mainly
interested in the subspace
\begin{math}
\mathcal{S}^A\otimes\mathcal{S}^B\subset\calH_p
\end{math}
of states which are \emph{totally symmetric} in each subensemble.
Within $\mathcal{S}_A$ or $\mathcal{S}_B$, the atoms behave like bosons. We
describe the atoms in subensemble $A$ by the bosonic operators $\hat{a}_s$
associated with the atomic levels $\ket{s}$. The atoms in $B$ are described by
similar operators $\hat{b}_s$.  Expressed in terms of these bosonic operators,
the Hamiltonian terms in Eq. (\ref{Hamiltonian:1}) read as
\begin{subequations}
\label{Hamiltonian:2}
\begin{align}
\label{Hamiltonian:2a}
\hatH_g & = g\left(\hata_2^\dag\hata_0+\hatb_2^\dag\hatb_0\right)\hatc
+ \mathrm{H.c.} \\
\label{Hamiltonian:2b}
\hatH_\Omega &= \Omega_a\hata_2^\dag\hata_1
+ \Omega_b\hatb_2^\dag\hatb_1 + \mathrm{H.c.}
\end{align}
\end{subequations}
Likewise, we rewrite the conservation of the total excitation number as
$\sum_{s=1}^{2}\hat{a}_{s}^{\dagger}\hat{a}_{s}
+\sum_{s=1}^{2}\hat{b}_{s}^{\dagger}\hat{b}_{s}
+\hat{c}^{\dagger}\hat{c} = p
$, and the constraints of having a fixed number of atoms in each subensemble as
$
\label{holonomy::eq:2}
\sum_{s=0}^{2}\hat{a}_{s}^{\dagger}\hat{a}_{s}=p$ and
$\sum_{s=0}^{2}\hat{b}_{s}^{\dagger}\hat{b}_{s} =n-p$.

Third, and most importantly, we bring into play the total occupation number in
the excited level $\hat{N}_2=\hat{a}_2^\dag\hat{a}_2+\hat{b}_2^\dag\hat{b}_2$,
with its associated even-odd parity operator:
\begin{equation}
\hat\Pi_2 := \exp(i\pi\hat{N}_2),
\end{equation}
which carries an interesting ``anti-symmetry'' \cite{endnote:2}:
\begin{equation}\label{anti_symmetry}
% \label{holonomy::eq:4}
\{\hat\Pi_2,\hat{H}\} = 0.
\end{equation}
This property follows from the fact that any atomic transition occurs through
the excited level $\ket{2}$.  The anti-symmetry implies that, in the parity
basis, the Hamiltonian is block-off-diagonal, and leads to one of our main
findings: \emph{the zero-energy subspace of $\calS_A\otimes\calS_B$ is always
  degenerate as long as $p>1$}.  We refer to the Supplemental Material
\cite{Supplemental} for the details of the general proof.
% Appendix~\ref{Appendixa}.

Within $\calS_A\otimes\calS_B$, we identify a zero-energy subspace which is decoherence-free by considering
% so that we can coherently manipulate quantum states in it.
% To this end, we note that
the limit of quantum Zeno dynamics ($g\to\infty$). Under this condition,
photon lekeage out of the cavity is completely suppressed within the subspace
$\mathcal{Z}\subset\calS_A\otimes\calS_B$ of zero-photon states, a so-called
\emph{quantum Zeno subspace} \cite{Facchi2002_PRL89-080401}. Since the
coupling of atoms with the electromagnetic field is mainly to the discrete
mode of the cavity, spontaneous decay from state $|2 \rangle$ is also strongly
suppressed \cite{Purcell46b, Beige2000_PRL85-1762, Facchi2002_PRL89-080401},
and $\cal Z$ can be considered as a decoherence-free subspace.  Within
$\cal Z$, $H_g=0$, and the Hamiltonian $\hatH$ $(=\hatH_\Omega)$ still bears
the anti-symmetry Eq.~(\ref{anti_symmetry}), hence the zero-energy subspace
$\cal D$ embedded in the Zeno subspace, $\cal D \subset \cal Z$, is always
degenerate for $p>1$. Besides being robust against photon decay, the states in
$\calD$ are \emph{dynamically irresponsive} to the external driving fields, as
$\hatH_\Omega=0$ within $\calD$. We call them ``dark states'' to distinguish
them from other zero-energy states outside $\calZ$.

In short, the subspace $\calD$ of dark states is our desired decoherence-free
subspace. Since the dark states are dynamically irresponsive, below we propose
to manipulate them by geometric means, that is, using non-Abelian geometric
phases (holonomies).  $\mathcal{D}$ is separated by a finite energy gap
$(\sim\Omega_{a/b})$ from the rest of the spectrum within $\calZ$, hence is
stable in quasi-adiabatic processes.

To be specific, from now on we will focus on the case of four atoms ($n=4$)
and two excitations ($p=2$).  Then, the quantum Zeno
subspace $\calZ$ consists of the following 6 basis states (excluding a state
which is completely decoupled from the rest):
\begin{align}
\ket{\zeta_1} &= \ket{0020,0} ,\;\;
\ket{\zeta_2}  = \ket{1010,0} ,\;\;
\ket{\zeta_3}  = \ket{2000,0} ,
\nonumber\\
\ket{\zeta_4} &= \frac{\ket{0002,0} -\ket{0101,0} +\ket{0200,0} }{\sqrt{3}},
\nonumber\\
\ket{\zeta_5} &= \frac{\ket{0110,0} -\sqrt{2} \ket{0011,0}
}{\sqrt{3}},
\nonumber\\
\ket{\zeta_6} &= \frac{\ket{1001,0} -\sqrt{2} \ket{1100,0} }{\sqrt{3}},
\end{align}
where $\ket{n_{a_1}n_{a_2}n_{b_1}n_{b_2},n_{c}}$ indicate the boson numbers.
Note that we have not specified the bosonic occupations of state $|0\rangle$, as they are fixed by the constraints
$n_{a_0} = p - n_{a_1} - n_{a_2}$
and $n_{b_0} = n-p - n_{b_1} - n_{b_2}$.  Within $\calZ$, the matrix
representation of the Hamiltonian in the specified basis is given by
\begin{equation}
\hatH \doteq
\begin{pmatrix}
0 & D^\dag \\
D & 0
\end{pmatrix}
\end{equation}
with the off-diagonal subblock
\begin{equation}
\label{effHamltonian}
\renewcommand\arraystretch{1.7}
D = 
\begin{pmatrix}
-\frac{2 \Omega_{b}}{\sqrt{3}} & \frac{\Omega_{a}}{\sqrt{3}} & 0 &
-\Omega_{b}^* \\
0 & \frac{\Omega_{b}}{\sqrt{3}} & -\frac{2 \Omega_{a}}{\sqrt{3}} &
-\Omega_{a}^*
\end{pmatrix} .
\end{equation}
It is clear that the dark-states subspace $\mathcal{D}$ is nothing but the
null space of $D$, hence is two-fold degenerate, in agreement with our general findings.  
Indeed, we find the following (unnormalized) basis states spanning $\calD$
\begin{equation}
\label{td1state}
\ket{D_1} = \ket{\zeta_1}\Omega_a^2
+ \ket{\zeta_2}2\Omega_a\Omega_b
+ \ket{\zeta_3}\Omega_b^2
\end{equation}
and
\begin{align}\label{td2state}
\ket{D_2}
= &\ket{\zeta_1}\sqrt{3}(\Omega_b^*)^2
\left(3|\Omega_a|^2+|\Omega_b|^2\right) \nonumber \\&{}
- \ket{\zeta_2}2\sqrt{3}\Omega_a^*\Omega_b^*
\left(|\Omega_a|^2+|\Omega_b|^2\right) \nonumber \\&{}
+ \ket{\zeta_3}\sqrt{3}(\Omega_a^*)^2
\left(|\Omega_a|^2+3|\Omega_b|^2\right) \nonumber \\&{}
- \ket{\zeta_4}
2\left(|\Omega_a|^4+4|\Omega_a\Omega_b|^2+|\Omega_b|^4\right) \,.
\end{align}

% \section{Holonomic Manipulation of Zero-Energy States}
\paragraph{Holonomic Manipulation.}
\label{Sec:Holonomy}

With a time-dependent Hamiltonian, the quantum state acquires not only
dynamical phases but also purely geometric phases, either Abelian
\cite{Berry84a} or non-Abelian \cite{Wilczek1984_PRL52-2111}.
Suppose that the Hamiltonian $\hatH(t)=\hatH(R_\mu(t))$ depending on slowly
varying control parameters $R_\mu(t)\in\mathbb{C}$ ($\mu=1,2,\dots$) maintains
a degenerate subspace of eigenstates $\ket{\eta_j(R_\mu(t))}$ ($j=1,2,\dots$)
at any instant $t$ of time. The adiabatic evolution of the states in the
subspace is governed by the unitary operator (up to a global phase factor)
\begin{equation}
\hatU(t,t')
= \sum_{ij}|\eta_i(R_\mu(t))\rangle U_{ij}(t,t')
\bra{\eta_j(R_\mu(t'))} .
\end{equation}
The adiabatic-evolution operator $\hatU$ depends only on the path
$\mathcal{C}$ in parameter space \cite{Wilczek1984_PRL52-2111}, and the
corresponding unitary matrix $U$ is given by
\begin{equation}
\label{Uexpression}
U(\mathcal{C})
= \mathcal{P}\exp\left(-\int_\mathcal{C}A^{\mu}dR_{\mu}\right) \,,
\end{equation}
where $\mathcal{P}$ denotes the path ordering and the matrix
\begin{equation}
\label{Aexpression}
A_{ij}^{\mu}=\langle\eta_{i}(R_\mu)|
\frac{\partial}{\partial R_{\mu}}|\eta_{j}(R_\mu)\rangle
\end{equation}
is the non-Abelian gauge potential describing the connection between the
instantaneous bases at different points in parameter space. We will denote
the non-Abelian holonomy interchangeably either by the operator $\hatU(\calC)$
or the matrix $U(\calC)$.

In our case, control parameters are the complex Rabi transition amplitudes,
$\Omega_{a}=\Omega\sin\theta\,e^{i\phi_{a}}$ and
$\Omega_{b}=\Omega\cos\theta\,e^{i\phi_{b}}$, and we modulate $\theta$ and
$\phi_\mu$ ($\mu=a,b$) in time. For most physical applications,
$\theta=\pi/4$ ($|\Omega_a|=|\Omega_b|$) is the most interesting
% system
configuration, and many adiabatic paths either (or both) start from
or end up with $\theta=\pi/4$. We find it convenient to split the path
$\calC$ into segments
\begin{math}
\calC = \calC_\theta + \calC_\phi + \calC_\theta' + \calC_\phi' + \dots \,.
\end{math}
In the amplitude-modulation segments $\calC_\theta(\theta_2,\theta_1)$, only
$\theta$ is varied from $\theta_1$ to $\theta_2$ keeping $\phi_\mu=0$. In the
phase-modulation segments $\calC_\phi(m_a,m_b;\theta)$, only $\phi_\mu$ are
modulated from 0 to $2\pi{m_\mu}$ ($m_\mu\in\mathbb{Z}$) with $\theta$ fixed.
% On this account, a phase modulation path, $\calC_\phi(m_a, m_b;\theta)$, is
% specified by the integer parameters $m_\mu$ and the fixed amplitude $\theta$.
% Depending on the context, we will drop either $m_\mu$ or $\theta$ to simplify
% notations.
% $\calC_\phi$ is always closed, but not necessarily so is
% $\calC_\theta$.

% \subsection{Phase Modulations}
% \paragraph{Phase Modulations}
% \label{Sec:specialclass}

The effects of amplitude and phase modulation are complementary: For
$\calC_\theta(\theta_1,\theta_0)$, the non-Abelian holonomy is given by
\cite{Supplemental}
\begin{equation}
\label{holonomy::eq:5}
U(\calC_\theta(\theta_1,\theta_0))=
\exp\left\{-i\left[c_y(\theta_1)-c_y(\theta_0)\right]\sigma_y\right\},
\end{equation}
with
\begin{math}
c_y(\theta) = -\arctan\sqrt{(19-5\cos4\theta)/6} \,,
\end{math}
where $\sigma^\mu$ are the Pauli matrices in the basis of~\eqref{td1state} and
\eqref{td2state}.  $U(\calC_\theta)$ thus describes a rotation around the
fixed $y$-axis, with only the angle depending on $\calC_\theta$.
On the other hand, $\calC_\phi(m_a,m_b;\theta)$ gives rise to
\cite{Supplemental}
\begin{eqnarray}
\label{uphi}
U(\calC_\phi) = \exp[ic_{x}\sigma_{x}+ic_{z}\sigma_{z}]
\end{eqnarray}
with the coefficients
\begin{align}
c_{x}= & \frac{2\sqrt{6}(m_a-m_b)\pi\sin2\theta\sin4\theta}
{(5-\cos4\theta)\sqrt{19-5\cos4\theta)}},
\nonumber\\
c_{z}= & \frac{(m_a+m_b)\pi(\cos8\theta-20\cos4\theta+51)}
{(5-\cos4\theta)(5\cos4\theta-19)}
\nonumber\\&
- \frac{(m_a-m_b)\pi(16\cos6\theta-96\cos2\theta)}
{(5-\cos4\theta)(5\cos4\theta-19)} \,.
\end{align}
$U(\calC_\phi)$ corresponds to a rotation around an axis in the $xz$-plane
with both axis and angle depending on $\calC_\phi$.

%%%%%%%%%%%%%%%%%%%%%%%%%%%%%%%%%%%%%%%%%%%%%%%%%%%%%%%%%%%%%%%%%%%%%%%%%%%%%%%
% \subsection{Universality}
\paragraph{Universality.}
\label{Sec:Universalitlity}
%%%%%%%%%%%%%%%%%%%%%%%%%%%%%%%%%%%%%%%%%%%%%%%%%%%%%%%%%%%%%%%%%%%%%%%%%%%%%%%

\begin{figure}
\centering
\includegraphics[width=60mm]{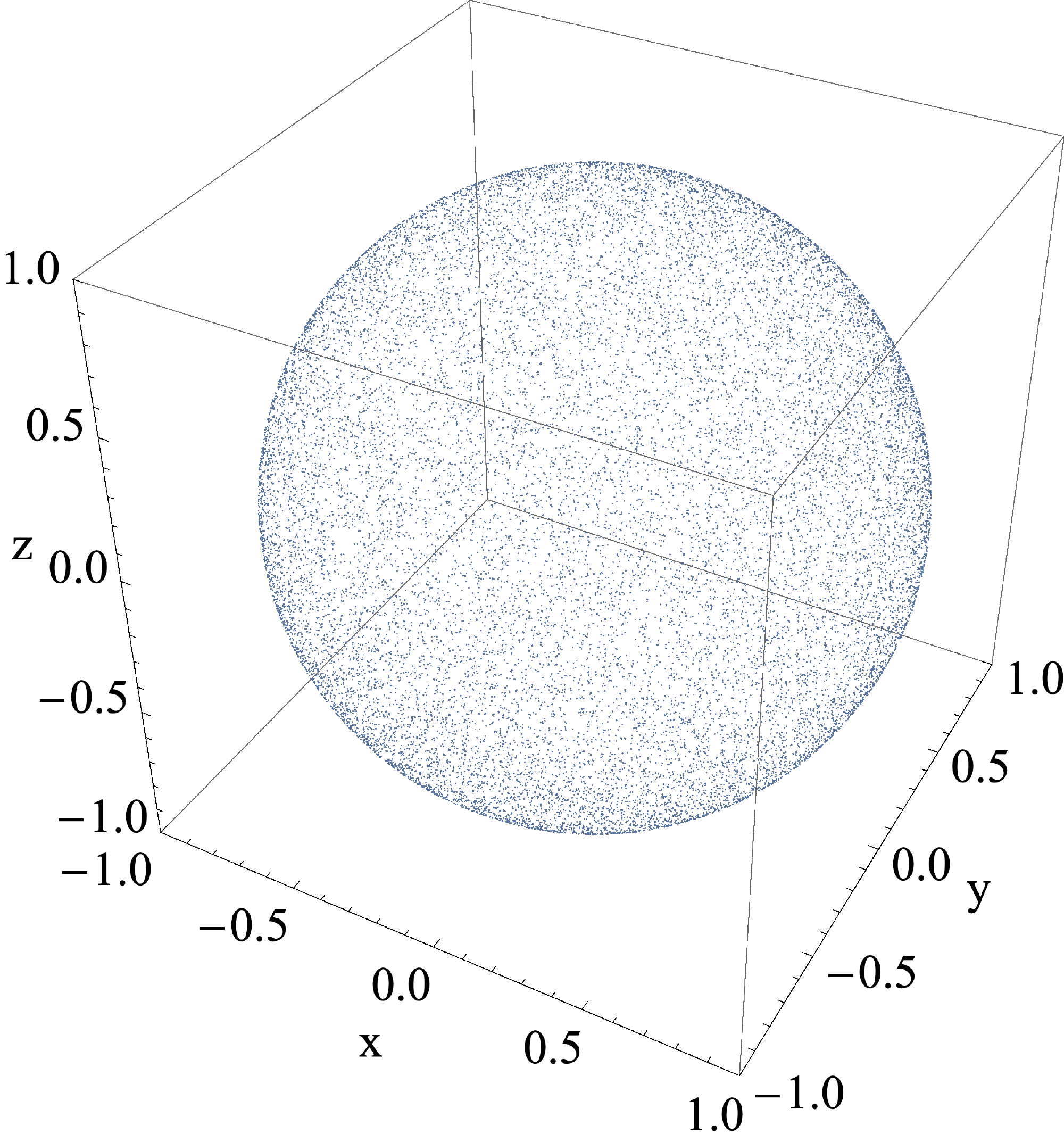}
\caption{Bloch sphere densely filled with states resulting from random
  sequences of two holonomies $\hatU_1$ and $\hatU_2$, applied to
  $\ket{D_1}$. We have chosen $\hatU_1=\hatU(\calC_\phi(1,0;\pi/6))$ and
  $\hatU_2=\hatU(\calC_\phi(0,-1;\pi/6))$.}
\label{fig-universality}
\end{figure}

Now we address the following question: Is it possible to implement an
arbitrary unitary transformation by combining $\hatU(\calC_\phi)$ and
$\hatU(\calC_\theta)$?  This is a non-trivial question, as the rotation axes
and angles of $\hatU(\calC_\phi)$ and $\hatU(\calC_\theta)$ of our concern are
not continuous.
However, we should recall that any two-dimensional unitary transformation can
be realized to arbitrary accuracy by combining two rotations around different
axes, if their angles are irrational multiples of $2\pi$. Therefore, given a
desired accuracy, we can implement any unitary transformation within
% the dark-states subspace
$\mathcal{D}$ by combining $\hatU(\calC_\phi)$ and $\hatU(\calC_\theta)$.
Alternatively, Fig.~\ref{fig-universality} demonstrates that two different
choices of $\hatU(\calC_\phi)$ are already sufficient. We have constructed
random sequences of $\hatU_1$ and $\hatU_2$ of varying lengths, with
$\hatU_1=\hatU(\calC_\phi(1,0;\pi/6))$ and
$\hatU_2=\hatU(\calC_\phi(0,-1;\pi/6))$, and applied them on $\ket{D_1}$.
Each point in Fig.~\ref{fig-universality} represents the resulting quantum
state. As seen, the states densely fill up the Bloch sphere.

For the purpose of demonstration, we further provide explicit adiabatic paths
that generate the elementary Pauli $X$ and $Z$. We consider the sequence:
\begin{multline}
\label{holonomy::eq:3}
\hatW(m_a,m_b;\theta_1) := \\{}
\hatU(\calC_\theta(\pi/4,\theta_1))
\hatU(\calC_\phi(m_a,m_b;\theta_1))
\hatU(\calC_\theta(\theta_1,\pi/4)) .
\end{multline}
Pauli $Z$ is extremely simple to implement, as it is identical to
$\hatW(1,0;\pi/4)$. 

While Pauli $X$ cannot be implemented exactly, the following procedure enables
an approximate implementation to arbitrary accuracy: First recall
from~Eq.~\eqref{holonomy::eq:5} that
\begin{math}
\hatU(\calC_\theta(\pi/4,\theta_1))=\hatU^\dag(\calC_\theta(\theta_1,\pi/4))
\end{math}
is a rotation around the $y$-axis. Therefore, just like $U(\calC_\phi)$, $\hatW$
is a rotation around an axis that is still in the $xz$-plane [see
Eq.~\eqref{uphi}].  One can numerically find a value $\theta_1^*$ such that 
$\hatW(m_a,m_b;\theta_1^*)$ is a rotation along the $x$-axis. The rotation angle is an irrational multiple of $2\pi$, thus $\hatW(m_a,m_b;\theta_1^*)$ does not yet realize the Pauli $X$. However, repeated applications of $\hatW(m_a,m_b;\theta_1^*)$
can reach any desired angle, say $\pi$, with arbitrary accuracy. This is
illustrated with $m_a=0$ and $m_b=1$ in Fig.~S1 of the Supplemental Material
\cite{Supplemental}.

% \section{Application}
\paragraph{Application.}
\label{Sec:application}

As an example of application of our scheme, we consider the generation of
symmetric Dicke states (unnormalized)
\begin{equation}
\label{Dickestate}
\ket{\Phi_n^p} = 
\sum_{i}\mathcal{P}_{i}
\ket{1}^{\otimes p}\otimes\ket{0}^{\otimes(n-p)}
\end{equation}
with directional ``angular momentum'' $(n/2-p)$, where
$\sum_{i}\mathcal{P}_{i}\cdots$ denotes the sum over distinct
permutations of all $n$ atoms.  Due to their rich entanglement
% in it especially for $p=n/2$
\cite{Toth07a,Toth09a}, Dicke states have attracted considerable interest as
valuable resources, e.g., for precision measurement
\cite{Apellaniz15a,Pezze2018_RMP90-035005,Holland93a}.

We first note that, regardless of parameters, there always exists a special
dark state which is immune to spontaneous decay from level $|2 \rangle$
% for a system of $n$ identical $\Lambda$-type atoms in a single-mode
% cavity with $p$ exciations, which is given by
(unnormalized):
\begin{equation}
\label{generaldarkstate}
\ket{E_n^p} =
\left[\Omega_a\hata_0^\dag - g\hatc\hata_1^\dag\right]^p
\left[\Omega_b\hatb_0^\dag - g\hatc\hatb_1^\dag\right]^{n-p}
(\hat c^\dag)^p\ket{\phantom{.}},
\end{equation}
where $\ket{\phantom{.}}$ denotes the vacuum state (no particle at all). 
A key observation is that, when $\Omega_a=\Omega_b$,
% the dark state
$\ket{E_n^p}$ is identical to the symmetric Dicke state $\ket{\Phi_n^p}$ in
the quantum Zeno limit.  More specifically, considering the example of
$n=4$ and $p=2$, $\ket{E_n^p}$ reads as (unnormalized)
\begin{equation}
\label{dickes}
\ket{E_4^2} =
|2000,0\rangle+2|1010,0\rangle+|0020,0\rangle
\end{equation}
in the bosonic notations or, equivalently:
\begin{align}
\label{dickes'}
\ket{E_4^2}
&=
 |1\rangle_1|1\rangle_2|0\rangle_3|0\rangle_4
+|1\rangle_1|0\rangle_2|1\rangle_3|0\rangle_4 \nonumber\\{}&
+|1\rangle_1|0\rangle_2|0\rangle_3|1\rangle_4
+|0\rangle_1|1\rangle_2|1\rangle_3|0\rangle_4 \nonumber\\{}&
+|0\rangle_1|1\rangle_2|0\rangle_3|1\rangle_4
+|0\rangle_1|0\rangle_2|1\rangle_3|1\rangle_4
\end{align} 
in the traditional representation.

We want to generate the dark state in~\eqref{dickes'} [or equivalently
\eqref{dickes}] starting from a product state
$|1\rangle_1|1\rangle_2|0\rangle_3|0\rangle_4$.  Note that both the initial
product state and the desired dark state in~\eqref{dickes'} belong to
$\calD$. The former for $\Omega_a=0$ and $\Omega_b=\Omega$ and the latter for
$\Omega_a=\Omega_b=\Omega/\sqrt{2}$. That is, the two states are adiabatically
connected and can be transformed into each other by the non-Abelian holonomy
discussed above. To this end, we slightly modify the sequence
in~\eqref{holonomy::eq:3} to
\begin{multline}
\label{holonomy::eq:1}
\hatW'(m_a,m_b;\theta_1) := \\
\hatU(\calC_\theta(\pi/4,\theta_1))
\hatU(\calC_\phi(m_a,m_b;\theta_1))
\hatU(\calC_\theta(\theta_1,0)) \,.
\end{multline}
In this sequence, $\theta$ starts from 0 and moves to $\theta_1$, then
$\phi_\mu$ make round trips from $0$ to integer multiples of $2\pi$, keeping
$\theta=\theta_1$, and finally $\theta$ moves from $\theta_1$ to $\pi/4$,
where the Hamiltonian becomes symmetric between subensembles $A$ and
$B$.
We find that the adiabatic paths specified by the parameters
$\{m_a,m_b\}=\{-24,1\}$, and $\theta_1=0.669$ in the
sequence \eqref{holonomy::eq:1} brings the initial product state to the
symmetric Dicke state in~\eqref{dickes'} with fidelity close to 1.

\begin{figure}
\centering
\includegraphics[width=80mm]{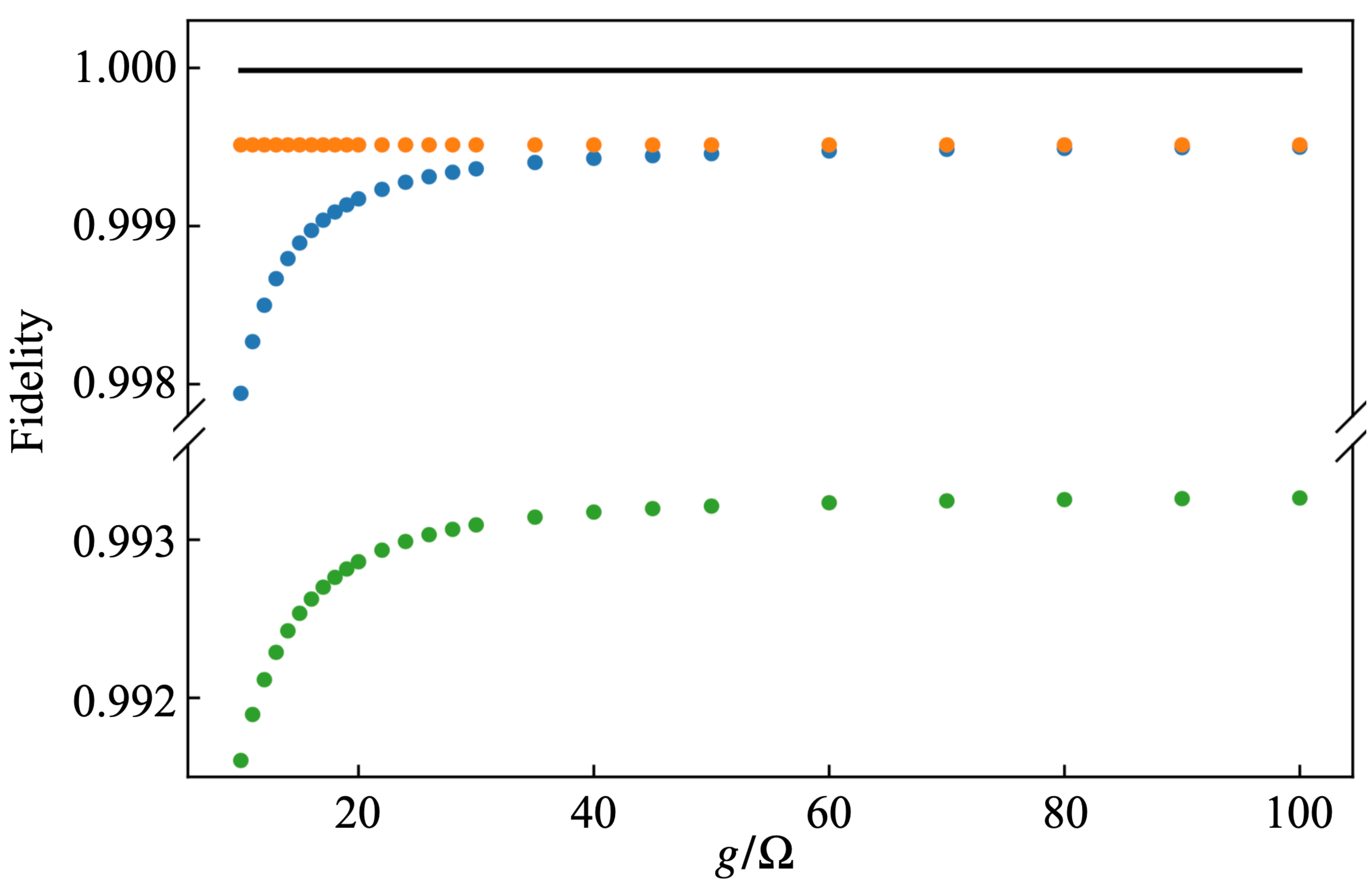}
\caption{Fidelity as a function of $g$ between the desired Dicke state and the
  state from the adiabatic evolution $\hatW'(-24,1;0.669)$ (black solid line).
  We also compute the fidelity by numerically solving the Schr\"odinger
  equation along the path
  $\calC_\theta(\pi/4,\theta_1) +\calC_\phi(-24,1;\theta_1)
  +\calC_\theta(\theta_1,0)$ within $\calZ$ (orange dots) and in the whole
  $\calS_A\otimes\calS_B$ (blue dots), where $\theta_1=0.669$.  For
  comparison, the fidelity of the state from the simulation along the path
  $\calC_\theta(\pi/4,0)$ is also shown with green dots.  The discrepancy
  between the orange dots and the black line is due to the finite simulation
  time, $T=8000/g$. \color{black}}
\label{fidelity}
\end{figure}

We also performed a time-dependent simulation with finite ramping time of the
parameters, by numerically solving the Schr\"odinger equation in the whole
space $\calS_A\otimes\calS_B$ (not restricted to $\calD$ or $\calZ$).  As
shown in Fig.~\ref{fidelity}, the simulation results agree very well with the
holonomic treatment, as long as $g/\Omega>10$.  This implies that, protected
by a finite energy gap in the spectrum, our holonomic method in the adiabatic
limit can be performed at realistic speeds.

We remark that it was previously proposed to achieve the same goal solely
based on the quantum Zeno dynamics \cite{Shao2010_EPL90-50003}, which
effectively corresponds to skipping the phase modulation sequence $\calC_\phi$
in Eq.~\eqref{holonomy::eq:1}. This approach also assumes a finite detuning in
the classical driving fields, which lifts the degeneracy in $\calD$ and
suppresses unwanted transitions to other states. In the absence of detuning,
however, transitions within $\calD$ degrade the fidelity of the final state
with the desired target state. The comparison in Fig.~\ref{fidelity} shows
that the phase modulation path $\calC_\phi$ plays a significant role to
readjust the state to the desired target state.

In conclusion, we have considered a system of atoms with $\Lambda$-type level
structure trapped in a single-mode cavity, and proposed a geometric scheme of
coherent manipulation on the subspace of zero-energy states within the quantum
Zeno subspace. These states inherit the decoherence-free nature of the quantum
Zeno subspace and feature a symmetry-protected generic degeneracy, fulfilling
all the conditions for a universal scheme of arbitrary unitary operations on
it.
Here we have taken a specific example with $n=4$ and $p=2$ for the purpose of
demonstration of the main idea. In principle, this method can be extended to
an arbitrary number of atoms and excitations.

%%%%%%%%%%%%%%%%%%%%%%%%%%%%%%%%%%%%%%%%%%%%%%%%%%%%%%%%%%%%%%%%
\acknowledgments

Y.-D. W.\ acknowledges support by the National Key R\&D Program of China under
Grant No. 2017YFA0304503, and the Peng Huanwu Theoretical Physics Renovation
Center under grant No. 12047503.
S.C.\ acknowledges support from the National Key Research and Development
Program of China (Grant No. 2016YFA0301200), NSFC (Grant No. 11974040), and
NSAF (Grant No. U1930402).
M.-S.C.\ has been supported by the National Research Function (NRF) of Korea
(Grant Nos. 2017R1E1A1A03070681 and 2018R1A4A1024157) and by the Ministry of
Education through the BK21 Four program.

%%%% References
%\bibliographystyle{apsrev4-1}
\bibliographystyle{apsrev4-2}
\bibliography{holonomy}

\begin{thebibliography}{48}%
\makeatletter
\providecommand \@ifxundefined [1]{%
 \@ifx{#1\undefined}
}%
\providecommand \@ifnum [1]{%
 \ifnum #1\expandafter \@firstoftwo
 \else \expandafter \@secondoftwo
 \fi
}%
\providecommand \@ifx [1]{%
 \ifx #1\expandafter \@firstoftwo
 \else \expandafter \@secondoftwo
 \fi
}%
\providecommand \natexlab [1]{#1}%
\providecommand \enquote  [1]{``#1''}%
\providecommand \bibnamefont  [1]{#1}%
\providecommand \bibfnamefont [1]{#1}%
\providecommand \citenamefont [1]{#1}%
\providecommand \href@noop [0]{\@secondoftwo}%
\providecommand \href [0]{\begingroup \@sanitize@url \@href}%
\providecommand \@href[1]{\@@startlink{#1}\@@href}%
\providecommand \@@href[1]{\endgroup#1\@@endlink}%
\providecommand \@sanitize@url [0]{\catcode `\\12\catcode `\$12\catcode
  `\&12\catcode `\#12\catcode `\^12\catcode `\_12\catcode `\%12\relax}%
\providecommand \@@startlink[1]{}%
\providecommand \@@endlink[0]{}%
\providecommand \url  [0]{\begingroup\@sanitize@url \@url }%
\providecommand \@url [1]{\endgroup\@href {#1}{\urlprefix }}%
\providecommand \urlprefix  [0]{URL }%
\providecommand \Eprint [0]{\href }%
\providecommand \doibase [0]{https://doi.org/}%
\providecommand \selectlanguage [0]{\@gobble}%
\providecommand \bibinfo  [0]{\@secondoftwo}%
\providecommand \bibfield  [0]{\@secondoftwo}%
\providecommand \translation [1]{[#1]}%
\providecommand \BibitemOpen [0]{}%
\providecommand \bibitemStop [0]{}%
\providecommand \bibitemNoStop [0]{.\EOS\space}%
\providecommand \EOS [0]{\spacefactor3000\relax}%
\providecommand \BibitemShut  [1]{\csname bibitem#1\endcsname}%
\let\auto@bib@innerbib\@empty
%</preamble>
\bibitem [{\citenamefont {Preskill}(2018)}]{Preskill18a}%
  \BibitemOpen
  \bibfield  {author} {\bibinfo {author} {\bibfnamefont {J.}~\bibnamefont
  {Preskill}},\ }\href {https://doi.org/10.22331/q-2018-08-06-79} {\bibfield
  {journal} {\bibinfo  {journal} {Quantum}\ }\textbf {\bibinfo {volume} {2}},\
  \bibinfo {pages} {79} (\bibinfo {year} {2018})}\BibitemShut {NoStop}%
\bibitem [{\citenamefont {Beige}\ \emph
  {et~al.}(2000{\natexlab{a}})\citenamefont {Beige}, \citenamefont {Braun},\
  and\ \citenamefont {Knight}}]{Beige2000_NJP2-22}%
  \BibitemOpen
  \bibfield  {author} {\bibinfo {author} {\bibfnamefont {A.}~\bibnamefont
  {Beige}}, \bibinfo {author} {\bibfnamefont {D.}~\bibnamefont {Braun}},\ and\
  \bibinfo {author} {\bibfnamefont {P.~L.}\ \bibnamefont {Knight}},\ }\href
  {https://doi.org/10.1088/1367-2630/2/1/322} {\bibfield  {journal} {\bibinfo
  {journal} {New Journal of Physics}\ }\textbf {\bibinfo {volume} {2}},\
  \bibinfo {pages} {22} (\bibinfo {year} {2000}{\natexlab{a}})}\BibitemShut
  {NoStop}%
\bibitem [{\citenamefont {Beige}\ \emph
  {et~al.}(2000{\natexlab{b}})\citenamefont {Beige}, \citenamefont {Braun},
  \citenamefont {Tregenna},\ and\ \citenamefont
  {Knight}}]{Beige2000_PRL85-1762}%
  \BibitemOpen
  \bibfield  {author} {\bibinfo {author} {\bibfnamefont {A.}~\bibnamefont
  {Beige}}, \bibinfo {author} {\bibfnamefont {D.}~\bibnamefont {Braun}},
  \bibinfo {author} {\bibfnamefont {B.}~\bibnamefont {Tregenna}},\ and\
  \bibinfo {author} {\bibfnamefont {P.~L.}\ \bibnamefont {Knight}},\ }\href
  {https://doi.org/10.1103/PhysRevLett.85.1762} {\bibfield  {journal} {\bibinfo
   {journal} {Phys. Rev. Lett.}\ }\textbf {\bibinfo {volume} {85}},\ \bibinfo
  {pages} {1762} (\bibinfo {year} {2000}{\natexlab{b}})}\BibitemShut {NoStop}%
\bibitem [{\citenamefont {Lidar}\ \emph {et~al.}(1998)\citenamefont {Lidar},
  \citenamefont {Chuang},\ and\ \citenamefont {Whaley}}]{Lidar1998_PRL81-2594}%
  \BibitemOpen
  \bibfield  {author} {\bibinfo {author} {\bibfnamefont {D.~A.}\ \bibnamefont
  {Lidar}}, \bibinfo {author} {\bibfnamefont {I.~L.}\ \bibnamefont {Chuang}},\
  and\ \bibinfo {author} {\bibfnamefont {K.~B.}\ \bibnamefont {Whaley}},\
  }\href {https://doi.org/10.1103/PhysRevLett.81.2594} {\bibfield  {journal}
  {\bibinfo  {journal} {Phys. Rev. Lett.}\ }\textbf {\bibinfo {volume} {81}},\
  \bibinfo {pages} {2594} (\bibinfo {year} {1998})}\BibitemShut {NoStop}%
\bibitem [{\citenamefont {Viola}\ and\ \citenamefont
  {Lloyd}(1998)}]{Viola1998_PRA58-2733}%
  \BibitemOpen
  \bibfield  {author} {\bibinfo {author} {\bibfnamefont {L.}~\bibnamefont
  {Viola}}\ and\ \bibinfo {author} {\bibfnamefont {S.}~\bibnamefont {Lloyd}},\
  }\href {https://doi.org/10.1103/PhysRevA.58.2733} {\bibfield  {journal}
  {\bibinfo  {journal} {Phys. Rev. A}\ }\textbf {\bibinfo {volume} {58}},\
  \bibinfo {pages} {2733} (\bibinfo {year} {1998})}\BibitemShut {NoStop}%
\bibitem [{\citenamefont {Vitali}\ and\ \citenamefont
  {Tombesi}(1999)}]{Vitali1999_PRA59-4178}%
  \BibitemOpen
  \bibfield  {author} {\bibinfo {author} {\bibfnamefont {D.}~\bibnamefont
  {Vitali}}\ and\ \bibinfo {author} {\bibfnamefont {P.}~\bibnamefont
  {Tombesi}},\ }\href {https://doi.org/10.1103/PhysRevA.59.4178} {\bibfield
  {journal} {\bibinfo  {journal} {Phys. Rev. A}\ }\textbf {\bibinfo {volume}
  {59}},\ \bibinfo {pages} {4178} (\bibinfo {year} {1999})}\BibitemShut
  {NoStop}%
\bibitem [{\citenamefont {Zanardi}(1999)}]{Zanardi1999_PLA258-77}%
  \BibitemOpen
  \bibfield  {author} {\bibinfo {author} {\bibfnamefont {P.}~\bibnamefont
  {Zanardi}},\ }\href
  {https://doi.org/https://doi.org/10.1016/S0375-9601(99)00365-5} {\bibfield
  {journal} {\bibinfo  {journal} {Physics Letters A}\ }\textbf {\bibinfo
  {volume} {258}},\ \bibinfo {pages} {77 } (\bibinfo {year}
  {1999})}\BibitemShut {NoStop}%
\bibitem [{\citenamefont {Shor}(1995)}]{Shor1995_PRA52-R2493}%
  \BibitemOpen
  \bibfield  {author} {\bibinfo {author} {\bibfnamefont {P.~W.}\ \bibnamefont
  {Shor}},\ }\href {https://doi.org/10.1103/PhysRevA.52.R2493} {\bibfield
  {journal} {\bibinfo  {journal} {Phys. Rev. A}\ }\textbf {\bibinfo {volume}
  {52}},\ \bibinfo {pages} {R2493} (\bibinfo {year} {1995})}\BibitemShut
  {NoStop}%
\bibitem [{\citenamefont {Steane}(1996)}]{Steane1996_PRL77-793}%
  \BibitemOpen
  \bibfield  {author} {\bibinfo {author} {\bibfnamefont {A.~M.}\ \bibnamefont
  {Steane}},\ }\href {https://doi.org/10.1103/PhysRevLett.77.793} {\bibfield
  {journal} {\bibinfo  {journal} {Phys. Rev. Lett.}\ }\textbf {\bibinfo
  {volume} {77}},\ \bibinfo {pages} {793} (\bibinfo {year} {1996})}\BibitemShut
  {NoStop}%
\bibitem [{\citenamefont {Plenio}\ \emph {et~al.}(1997)\citenamefont {Plenio},
  \citenamefont {Vedral},\ and\ \citenamefont {Knight}}]{Plenio1997_PRA55-67}%
  \BibitemOpen
  \bibfield  {author} {\bibinfo {author} {\bibfnamefont {M.~B.}\ \bibnamefont
  {Plenio}}, \bibinfo {author} {\bibfnamefont {V.}~\bibnamefont {Vedral}},\
  and\ \bibinfo {author} {\bibfnamefont {P.~L.}\ \bibnamefont {Knight}},\
  }\href {https://doi.org/10.1103/PhysRevA.55.67} {\bibfield  {journal}
  {\bibinfo  {journal} {Phys. Rev. A}\ }\textbf {\bibinfo {volume} {55}},\
  \bibinfo {pages} {67} (\bibinfo {year} {1997})}\BibitemShut {NoStop}%
\bibitem [{\citenamefont {Zanardi}\ and\ \citenamefont
  {Rasetti}(1999)}]{Zanardi1999_PLA264-94}%
  \BibitemOpen
  \bibfield  {author} {\bibinfo {author} {\bibfnamefont {P.}~\bibnamefont
  {Zanardi}}\ and\ \bibinfo {author} {\bibfnamefont {M.}~\bibnamefont
  {Rasetti}},\ }\href
  {https://doi.org/https://doi.org/10.1016/S0375-9601(99)00803-8} {\bibfield
  {journal} {\bibinfo  {journal} {Physics Letters A}\ }\textbf {\bibinfo
  {volume} {264}},\ \bibinfo {pages} {94 } (\bibinfo {year}
  {1999})}\BibitemShut {NoStop}%
\bibitem [{\citenamefont {Jones}\ \emph {et~al.}(2000)\citenamefont {Jones},
  \citenamefont {Vedral}, \citenamefont {Ekert},\ and\ \citenamefont
  {Castagnoli}}]{Jones2000_Nature403-869}%
  \BibitemOpen
  \bibfield  {author} {\bibinfo {author} {\bibfnamefont {J.~A.}\ \bibnamefont
  {Jones}}, \bibinfo {author} {\bibfnamefont {V.}~\bibnamefont {Vedral}},
  \bibinfo {author} {\bibfnamefont {A.}~\bibnamefont {Ekert}},\ and\ \bibinfo
  {author} {\bibfnamefont {G.}~\bibnamefont {Castagnoli}},\ }\href
  {https://doi.org/10.1038/35002528} {\bibfield  {journal} {\bibinfo  {journal}
  {Nature}\ }\textbf {\bibinfo {volume} {403}},\ \bibinfo {pages} {869}
  (\bibinfo {year} {2000})}\BibitemShut {NoStop}%
\bibitem [{\citenamefont {Duan}\ \emph {et~al.}(2001)\citenamefont {Duan},
  \citenamefont {Cirac},\ and\ \citenamefont
  {Zoller}}]{Duan2001_Science292-1695}%
  \BibitemOpen
  \bibfield  {author} {\bibinfo {author} {\bibfnamefont {L.-M.}\ \bibnamefont
  {Duan}}, \bibinfo {author} {\bibfnamefont {J.~I.}\ \bibnamefont {Cirac}},\
  and\ \bibinfo {author} {\bibfnamefont {P.}~\bibnamefont {Zoller}},\ }\href
  {https://doi.org/10.1126/science.1058835} {\bibfield  {journal} {\bibinfo
  {journal} {Science}\ }\textbf {\bibinfo {volume} {292}},\ \bibinfo {pages}
  {1695} (\bibinfo {year} {2001})}\BibitemShut {NoStop}%
\bibitem [{\citenamefont {Sj{\"o}qvist}\ \emph {et~al.}(2012)\citenamefont
  {Sj{\"o}qvist}, \citenamefont {Tong}, \citenamefont {Mauritz~Andersson},
  \citenamefont {Hessmo}, \citenamefont {Johansson},\ and\ \citenamefont
  {Singh}}]{Sjoqvist12a}%
  \BibitemOpen
  \bibfield  {author} {\bibinfo {author} {\bibfnamefont {E.}~\bibnamefont
  {Sj{\"o}qvist}}, \bibinfo {author} {\bibfnamefont {D.~M.}\ \bibnamefont
  {Tong}}, \bibinfo {author} {\bibfnamefont {L.}~\bibnamefont
  {Mauritz~Andersson}}, \bibinfo {author} {\bibfnamefont {B.}~\bibnamefont
  {Hessmo}}, \bibinfo {author} {\bibfnamefont {M.}~\bibnamefont {Johansson}},\
  and\ \bibinfo {author} {\bibfnamefont {K.}~\bibnamefont {Singh}},\ }\href
  {https://doi.org/10.1088/1367-2630/14/10/103035} {\bibfield  {journal}
  {\bibinfo  {journal} {New Journal of Physics}\ }\textbf {\bibinfo {volume}
  {14}},\ \bibinfo {pages} {103035} (\bibinfo {year} {2012})}\BibitemShut
  {NoStop}%
\bibitem [{\citenamefont {Toyoda}\ \emph {et~al.}(2013)\citenamefont {Toyoda},
  \citenamefont {Uchida}, \citenamefont {Noguchi}, \citenamefont {Haze},\ and\
  \citenamefont {Urabe}}]{Toyoda2013_PRA87-052307}%
  \BibitemOpen
  \bibfield  {author} {\bibinfo {author} {\bibfnamefont {K.}~\bibnamefont
  {Toyoda}}, \bibinfo {author} {\bibfnamefont {K.}~\bibnamefont {Uchida}},
  \bibinfo {author} {\bibfnamefont {A.}~\bibnamefont {Noguchi}}, \bibinfo
  {author} {\bibfnamefont {S.}~\bibnamefont {Haze}},\ and\ \bibinfo {author}
  {\bibfnamefont {S.}~\bibnamefont {Urabe}},\ }\href
  {https://doi.org/10.1103/PhysRevA.87.052307} {\bibfield  {journal} {\bibinfo
  {journal} {Phys. Rev. A}\ }\textbf {\bibinfo {volume} {87}},\ \bibinfo
  {pages} {052307} (\bibinfo {year} {2013})}\BibitemShut {NoStop}%
\bibitem [{\citenamefont {Feng}\ \emph {et~al.}(2013)\citenamefont {Feng},
  \citenamefont {Xu},\ and\ \citenamefont {Long}}]{Feng2013_PRL110-190501}%
  \BibitemOpen
  \bibfield  {author} {\bibinfo {author} {\bibfnamefont {G.}~\bibnamefont
  {Feng}}, \bibinfo {author} {\bibfnamefont {G.}~\bibnamefont {Xu}},\ and\
  \bibinfo {author} {\bibfnamefont {G.}~\bibnamefont {Long}},\ }\href
  {https://doi.org/10.1103/PhysRevLett.110.190501} {\bibfield  {journal}
  {\bibinfo  {journal} {Phys. Rev. Lett.}\ }\textbf {\bibinfo {volume} {110}},\
  \bibinfo {pages} {190501} (\bibinfo {year} {2013})}\BibitemShut {NoStop}%
\bibitem [{\citenamefont {Abdumalikov~Jr}\ \emph {et~al.}(2013)\citenamefont
  {Abdumalikov~Jr}, \citenamefont {Fink}, \citenamefont {Juliusson},
  \citenamefont {Pechal}, \citenamefont {Berger}, \citenamefont {Wallraff},\
  and\ \citenamefont {Filipp}}]{Abdumalikov2013_Nature496-482}%
  \BibitemOpen
  \bibfield  {author} {\bibinfo {author} {\bibfnamefont {A.~A.}\ \bibnamefont
  {Abdumalikov~Jr}}, \bibinfo {author} {\bibfnamefont {J.~M.}\ \bibnamefont
  {Fink}}, \bibinfo {author} {\bibfnamefont {K.}~\bibnamefont {Juliusson}},
  \bibinfo {author} {\bibfnamefont {M.}~\bibnamefont {Pechal}}, \bibinfo
  {author} {\bibfnamefont {S.}~\bibnamefont {Berger}}, \bibinfo {author}
  {\bibfnamefont {A.}~\bibnamefont {Wallraff}},\ and\ \bibinfo {author}
  {\bibfnamefont {S.}~\bibnamefont {Filipp}},\ }\href
  {https://doi.org/10.1038/nature12010} {\bibfield  {journal} {\bibinfo
  {journal} {Nature}\ }\textbf {\bibinfo {volume} {496}},\ \bibinfo {pages}
  {482} (\bibinfo {year} {2013})}\BibitemShut {NoStop}%
\bibitem [{\citenamefont {Zu}\ \emph {et~al.}(2014)\citenamefont {Zu},
  \citenamefont {Wang}, \citenamefont {He}, \citenamefont {Zhang},
  \citenamefont {Dai}, \citenamefont {Wang},\ and\ \citenamefont
  {Duan}}]{Zu14a}%
  \BibitemOpen
  \bibfield  {author} {\bibinfo {author} {\bibfnamefont {C.}~\bibnamefont
  {Zu}}, \bibinfo {author} {\bibfnamefont {W.-B.}\ \bibnamefont {Wang}},
  \bibinfo {author} {\bibfnamefont {L.}~\bibnamefont {He}}, \bibinfo {author}
  {\bibfnamefont {W.-G.}\ \bibnamefont {Zhang}}, \bibinfo {author}
  {\bibfnamefont {C.-Y.}\ \bibnamefont {Dai}}, \bibinfo {author} {\bibfnamefont
  {F.}~\bibnamefont {Wang}},\ and\ \bibinfo {author} {\bibfnamefont {L.-M.}\
  \bibnamefont {Duan}},\ }\href {https://doi.org/10.1038/nature13729}
  {\bibfield  {journal} {\bibinfo  {journal} {Nature}\ }\textbf {\bibinfo
  {volume} {514}},\ \bibinfo {pages} {72} (\bibinfo {year} {2014})}\BibitemShut
  {NoStop}%
\bibitem [{\citenamefont {Xu}\ \emph {et~al.}(2018)\citenamefont {Xu},
  \citenamefont {Cai}, \citenamefont {Ma}, \citenamefont {Mu}, \citenamefont
  {Hu}, \citenamefont {Chen}, \citenamefont {Wang}, \citenamefont {Song},
  \citenamefont {Xue}, \citenamefont {Yin},\ and\ \citenamefont
  {Sun}}]{Xu2018_PRL121-110501}%
  \BibitemOpen
  \bibfield  {author} {\bibinfo {author} {\bibfnamefont {Y.}~\bibnamefont
  {Xu}}, \bibinfo {author} {\bibfnamefont {W.}~\bibnamefont {Cai}}, \bibinfo
  {author} {\bibfnamefont {Y.}~\bibnamefont {Ma}}, \bibinfo {author}
  {\bibfnamefont {X.}~\bibnamefont {Mu}}, \bibinfo {author} {\bibfnamefont
  {L.}~\bibnamefont {Hu}}, \bibinfo {author} {\bibfnamefont {T.}~\bibnamefont
  {Chen}}, \bibinfo {author} {\bibfnamefont {H.}~\bibnamefont {Wang}}, \bibinfo
  {author} {\bibfnamefont {Y.~P.}\ \bibnamefont {Song}}, \bibinfo {author}
  {\bibfnamefont {Z.-Y.}\ \bibnamefont {Xue}}, \bibinfo {author} {\bibfnamefont
  {Z.-q.}\ \bibnamefont {Yin}},\ and\ \bibinfo {author} {\bibfnamefont
  {L.}~\bibnamefont {Sun}},\ }\href
  {https://doi.org/10.1103/PhysRevLett.121.110501} {\bibfield  {journal}
  {\bibinfo  {journal} {Phys. Rev. Lett.}\ }\textbf {\bibinfo {volume} {121}},\
  \bibinfo {pages} {110501} (\bibinfo {year} {2018})}\BibitemShut {NoStop}%
\bibitem [{\citenamefont {Huang}\ \emph {et~al.}(2019)\citenamefont {Huang},
  \citenamefont {Wu}, \citenamefont {Wang}, \citenamefont {Hou}, \citenamefont
  {Wang}, \citenamefont {Zhang}, \citenamefont {Lian}, \citenamefont {Liu},
  \citenamefont {Wang}, \citenamefont {Zhang}, \citenamefont {He},
  \citenamefont {Chang}, \citenamefont {Xu},\ and\ \citenamefont
  {Duan}}]{Duan2019_PRL122-010503}%
  \BibitemOpen
  \bibfield  {author} {\bibinfo {author} {\bibfnamefont {Y.-Y.}\ \bibnamefont
  {Huang}}, \bibinfo {author} {\bibfnamefont {Y.-K.}\ \bibnamefont {Wu}},
  \bibinfo {author} {\bibfnamefont {F.}~\bibnamefont {Wang}}, \bibinfo {author}
  {\bibfnamefont {P.-Y.}\ \bibnamefont {Hou}}, \bibinfo {author} {\bibfnamefont
  {W.-B.}\ \bibnamefont {Wang}}, \bibinfo {author} {\bibfnamefont {W.-G.}\
  \bibnamefont {Zhang}}, \bibinfo {author} {\bibfnamefont {W.-Q.}\ \bibnamefont
  {Lian}}, \bibinfo {author} {\bibfnamefont {Y.-Q.}\ \bibnamefont {Liu}},
  \bibinfo {author} {\bibfnamefont {H.-Y.}\ \bibnamefont {Wang}}, \bibinfo
  {author} {\bibfnamefont {H.-Y.}\ \bibnamefont {Zhang}}, \bibinfo {author}
  {\bibfnamefont {L.}~\bibnamefont {He}}, \bibinfo {author} {\bibfnamefont
  {X.-Y.}\ \bibnamefont {Chang}}, \bibinfo {author} {\bibfnamefont
  {Y.}~\bibnamefont {Xu}},\ and\ \bibinfo {author} {\bibfnamefont {L.-M.}\
  \bibnamefont {Duan}},\ }\href
  {https://doi.org/10.1103/PhysRevLett.122.010503} {\bibfield  {journal}
  {\bibinfo  {journal} {Phys. Rev. Lett.}\ }\textbf {\bibinfo {volume} {122}},\
  \bibinfo {pages} {010503} (\bibinfo {year} {2019})}\BibitemShut {NoStop}%
\bibitem [{\citenamefont {Freedman}(1998)}]{Freedman1998_PNAS95-98}%
  \BibitemOpen
  \bibfield  {author} {\bibinfo {author} {\bibfnamefont {M.~H.}\ \bibnamefont
  {Freedman}},\ }\href {https://doi.org/10.1073/pnas.95.1.98} {\bibfield
  {journal} {\bibinfo  {journal} {Proceedings of the National Academy of
  Sciences}\ }\textbf {\bibinfo {volume} {95}},\ \bibinfo {pages} {98}
  (\bibinfo {year} {1998})}\BibitemShut {NoStop}%
\bibitem [{\citenamefont {Kitaev}(2003)}]{Kitaev2003_AnnalsofPhysics303-2}%
  \BibitemOpen
  \bibfield  {author} {\bibinfo {author} {\bibfnamefont {A.}~\bibnamefont
  {Kitaev}},\ }\href
  {https://doi.org/https://doi.org/10.1016/S0003-4916(02)00018-0} {\bibfield
  {journal} {\bibinfo  {journal} {Annals of Physics}\ }\textbf {\bibinfo
  {volume} {303}},\ \bibinfo {pages} {2 } (\bibinfo {year} {2003})}\BibitemShut
  {NoStop}%
\bibitem [{\citenamefont {Nayak}\ \emph {et~al.}(2008)\citenamefont {Nayak},
  \citenamefont {Simon}, \citenamefont {Stern}, \citenamefont {Freedman},\ and\
  \citenamefont {Sarma}}]{Nayak08a}%
  \BibitemOpen
  \bibfield  {author} {\bibinfo {author} {\bibfnamefont {C.}~\bibnamefont
  {Nayak}}, \bibinfo {author} {\bibfnamefont {S.~H.}\ \bibnamefont {Simon}},
  \bibinfo {author} {\bibfnamefont {A.}~\bibnamefont {Stern}}, \bibinfo
  {author} {\bibfnamefont {M.}~\bibnamefont {Freedman}},\ and\ \bibinfo
  {author} {\bibfnamefont {S.~D.}\ \bibnamefont {Sarma}},\ }\href
  {https://doi.org/10.1103/RevModPhys.80.1083} {\bibfield  {journal} {\bibinfo
  {journal} {Rev. Mod. Phys.}\ }\textbf {\bibinfo {volume} {80}},\ \bibinfo
  {pages} {1083} (\bibinfo {year} {2008})}\BibitemShut {NoStop}%
\bibitem [{\citenamefont {Mourik}\ \emph {et~al.}(2012)\citenamefont {Mourik},
  \citenamefont {Zuo}, \citenamefont {Frolov}, \citenamefont {Plissard},
  \citenamefont {Bakkers},\ and\ \citenamefont {Kouwenhoven}}]{Mourik12a}%
  \BibitemOpen
  \bibfield  {author} {\bibinfo {author} {\bibfnamefont {V.}~\bibnamefont
  {Mourik}}, \bibinfo {author} {\bibfnamefont {K.}~\bibnamefont {Zuo}},
  \bibinfo {author} {\bibfnamefont {S.~M.}\ \bibnamefont {Frolov}}, \bibinfo
  {author} {\bibfnamefont {S.~R.}\ \bibnamefont {Plissard}}, \bibinfo {author}
  {\bibfnamefont {E.~P. A.~M.}\ \bibnamefont {Bakkers}},\ and\ \bibinfo
  {author} {\bibfnamefont {L.~P.}\ \bibnamefont {Kouwenhoven}},\ }\href
  {https://doi.org/10.1126/science.1222360} {\bibfield  {journal} {\bibinfo
  {journal} {Science}\ }\textbf {\bibinfo {volume} {336}},\ \bibinfo {pages}
  {1003} (\bibinfo {year} {2012})}\BibitemShut {NoStop}%
\bibitem [{\citenamefont {Deng}\ \emph {et~al.}(2012)\citenamefont {Deng},
  \citenamefont {Yu}, \citenamefont {Huang}, \citenamefont {Larsson},
  \citenamefont {Caroff},\ and\ \citenamefont {Xu}}]{Deng12a}%
  \BibitemOpen
  \bibfield  {author} {\bibinfo {author} {\bibfnamefont {M.~T.}\ \bibnamefont
  {Deng}}, \bibinfo {author} {\bibfnamefont {C.~L.}\ \bibnamefont {Yu}},
  \bibinfo {author} {\bibfnamefont {G.~Y.}\ \bibnamefont {Huang}}, \bibinfo
  {author} {\bibfnamefont {M.}~\bibnamefont {Larsson}}, \bibinfo {author}
  {\bibfnamefont {P.}~\bibnamefont {Caroff}},\ and\ \bibinfo {author}
  {\bibfnamefont {H.~Q.}\ \bibnamefont {Xu}},\ }\href
  {https://doi.org/10.1021/nl303758w} {\bibfield  {journal} {\bibinfo
  {journal} {Nano Letters}\ }\textbf {\bibinfo {volume} {12}},\ \bibinfo
  {pages} {6414} (\bibinfo {year} {2012})}\BibitemShut {NoStop}%
\bibitem [{\citenamefont {Das}\ \emph {et~al.}(2012)\citenamefont {Das},
  \citenamefont {Ronen}, \citenamefont {Most}, \citenamefont {Oreg},
  \citenamefont {Heiblum},\ and\ \citenamefont {Shtrikman}}]{Das12a}%
  \BibitemOpen
  \bibfield  {author} {\bibinfo {author} {\bibfnamefont {A.}~\bibnamefont
  {Das}}, \bibinfo {author} {\bibfnamefont {Y.}~\bibnamefont {Ronen}}, \bibinfo
  {author} {\bibfnamefont {Y.}~\bibnamefont {Most}}, \bibinfo {author}
  {\bibfnamefont {Y.}~\bibnamefont {Oreg}}, \bibinfo {author} {\bibfnamefont
  {M.}~\bibnamefont {Heiblum}},\ and\ \bibinfo {author} {\bibfnamefont
  {H.}~\bibnamefont {Shtrikman}},\ }\href {https://doi.org/10.1038/nphys2479}
  {\bibfield  {journal} {\bibinfo  {journal} {Nat Phys}\ }\textbf {\bibinfo
  {volume} {8}},\ \bibinfo {pages} {887} (\bibinfo {year} {2012})}\BibitemShut
  {NoStop}%
\bibitem [{\citenamefont {Nadj-Perge}\ \emph {et~al.}(2014)\citenamefont
  {Nadj-Perge}, \citenamefont {Drozdov}, \citenamefont {Li}, \citenamefont
  {Chen}, \citenamefont {Jeon}, \citenamefont {Seo}, \citenamefont {MacDonald},
  \citenamefont {Bernevig},\ and\ \citenamefont {Yazdani}}]{Nadj-Perge14a}%
  \BibitemOpen
  \bibfield  {author} {\bibinfo {author} {\bibfnamefont {S.}~\bibnamefont
  {Nadj-Perge}}, \bibinfo {author} {\bibfnamefont {I.~K.}\ \bibnamefont
  {Drozdov}}, \bibinfo {author} {\bibfnamefont {J.}~\bibnamefont {Li}},
  \bibinfo {author} {\bibfnamefont {H.}~\bibnamefont {Chen}}, \bibinfo {author}
  {\bibfnamefont {S.}~\bibnamefont {Jeon}}, \bibinfo {author} {\bibfnamefont
  {J.}~\bibnamefont {Seo}}, \bibinfo {author} {\bibfnamefont {A.~H.}\
  \bibnamefont {MacDonald}}, \bibinfo {author} {\bibfnamefont {B.~A.}\
  \bibnamefont {Bernevig}},\ and\ \bibinfo {author} {\bibfnamefont
  {A.}~\bibnamefont {Yazdani}},\ }\href
  {https://doi.org/10.1126/science.1259327} {\bibfield  {journal} {\bibinfo
  {journal} {Science}\ }\textbf {\bibinfo {volume} {346}},\ \bibinfo {pages}
  {602} (\bibinfo {year} {2014})}\BibitemShut {NoStop}%
\bibitem [{\citenamefont {Bartolomei}\ \emph {et~al.}(2020)\citenamefont
  {Bartolomei}, \citenamefont {Kumar}, \citenamefont {Bisognin}, \citenamefont
  {Marguerite}, \citenamefont {Berroir}, \citenamefont {Bocquillon},
  \citenamefont {Pla{\c{c}}ais}, \citenamefont {Cavanna}, \citenamefont {Dong},
  \citenamefont {Gennser}, \citenamefont {Jin},\ and\ \citenamefont
  {F{\`{e}}ve}}]{Bartolomei20a}%
  \BibitemOpen
  \bibfield  {author} {\bibinfo {author} {\bibfnamefont {H.}~\bibnamefont
  {Bartolomei}}, \bibinfo {author} {\bibfnamefont {M.}~\bibnamefont {Kumar}},
  \bibinfo {author} {\bibfnamefont {R.}~\bibnamefont {Bisognin}}, \bibinfo
  {author} {\bibfnamefont {A.}~\bibnamefont {Marguerite}}, \bibinfo {author}
  {\bibfnamefont {J.-M.}\ \bibnamefont {Berroir}}, \bibinfo {author}
  {\bibfnamefont {E.}~\bibnamefont {Bocquillon}}, \bibinfo {author}
  {\bibfnamefont {B.}~\bibnamefont {Pla{\c{c}}ais}}, \bibinfo {author}
  {\bibfnamefont {A.}~\bibnamefont {Cavanna}}, \bibinfo {author} {\bibfnamefont
  {Q.}~\bibnamefont {Dong}}, \bibinfo {author} {\bibfnamefont {U.}~\bibnamefont
  {Gennser}}, \bibinfo {author} {\bibfnamefont {Y.}~\bibnamefont {Jin}},\ and\
  \bibinfo {author} {\bibfnamefont {G.}~\bibnamefont {F{\`{e}}ve}},\ }\href
  {https://doi.org/10.1126/science.aaz5601} {\bibfield  {journal} {\bibinfo
  {journal} {Science}\ }\textbf {\bibinfo {volume} {368}},\ \bibinfo {pages}
  {173} (\bibinfo {year} {2020})}\BibitemShut {NoStop}%
\bibitem [{end({\natexlab{a}})}]{endnote:1}%
  \BibitemOpen
  \href@noop {} {} ({\natexlab{a}}),\ \bibinfo {note} {our scheme is not
  limited to real atoms but also applies to other systems as long as the
  contituent elements have the $\Lambda$-type level structure. A common example
  is the nitrogen-vacancy center in diamonds; see
  Refs.~\cite{Zu14a,Yang2010_NJP12-113039}.}\BibitemShut {Stop}%
\bibitem [{\citenamefont {Aharonov}\ and\ \citenamefont
  {Anandan}(1987)}]{Aharonov87a}%
  \BibitemOpen
  \bibfield  {author} {\bibinfo {author} {\bibfnamefont {Y.}~\bibnamefont
  {Aharonov}}\ and\ \bibinfo {author} {\bibfnamefont {J.}~\bibnamefont
  {Anandan}},\ }\href {https://doi.org/10.1103/PhysRevLett.58.1593} {\bibfield
  {journal} {\bibinfo  {journal} {Phys. Rev. Lett.}\ }\textbf {\bibinfo
  {volume} {58}},\ \bibinfo {pages} {1593} (\bibinfo {year}
  {1987})}\BibitemShut {NoStop}%
\bibitem [{\citenamefont {Facchi}\ and\ \citenamefont
  {Pascazio}(2002)}]{Facchi2002_PRL89-080401}%
  \BibitemOpen
  \bibfield  {author} {\bibinfo {author} {\bibfnamefont {P.}~\bibnamefont
  {Facchi}}\ and\ \bibinfo {author} {\bibfnamefont {S.}~\bibnamefont
  {Pascazio}},\ }\href {https://doi.org/10.1103/PhysRevLett.89.080401}
  {\bibfield  {journal} {\bibinfo  {journal} {Phys. Rev. Lett.}\ }\textbf
  {\bibinfo {volume} {89}},\ \bibinfo {pages} {080401} (\bibinfo {year}
  {2002})}\BibitemShut {NoStop}%
\bibitem [{\citenamefont {Wu}\ and\ \citenamefont
  {Su}(2019)}]{Wu2019_JPA52-335301}%
  \BibitemOpen
  \bibfield  {author} {\bibinfo {author} {\bibfnamefont {J.~L.}\ \bibnamefont
  {Wu}}\ and\ \bibinfo {author} {\bibfnamefont {S.~L.}\ \bibnamefont {Su}},\
  }\href {https://doi.org/10.1088/1751-8121/ab2a92} {\bibfield  {journal}
  {\bibinfo  {journal} {Journal of Physics A: Mathematical and Theoretical}\
  }\textbf {\bibinfo {volume} {52}},\ \bibinfo {pages} {335301} (\bibinfo
  {year} {2019})}\BibitemShut {NoStop}%
\bibitem [{\citenamefont {Song}\ \emph {et~al.}(2016)\citenamefont {Song},
  \citenamefont {Zhang}, \citenamefont {Ai}, \citenamefont {Qiu},\ and\
  \citenamefont {Deng}}]{Song2016_NJP18-023001}%
  \BibitemOpen
  \bibfield  {author} {\bibinfo {author} {\bibfnamefont {X.-K.}\ \bibnamefont
  {Song}}, \bibinfo {author} {\bibfnamefont {H.}~\bibnamefont {Zhang}},
  \bibinfo {author} {\bibfnamefont {Q.}~\bibnamefont {Ai}}, \bibinfo {author}
  {\bibfnamefont {J.}~\bibnamefont {Qiu}},\ and\ \bibinfo {author}
  {\bibfnamefont {F.-G.}\ \bibnamefont {Deng}},\ }\href
  {https://doi.org/10.1088/1367-2630/18/2/023001} {\bibfield  {journal}
  {\bibinfo  {journal} {New Journal of Physics}\ }\textbf {\bibinfo {volume}
  {18}},\ \bibinfo {pages} {023001} (\bibinfo {year} {2016})}\BibitemShut
  {NoStop}%
\bibitem [{\citenamefont {Mousolou}\ and\ \citenamefont
  {Sj{\"o}qvist}(2018)}]{Mousolou2018_JPA51-475303}%
  \BibitemOpen
  \bibfield  {author} {\bibinfo {author} {\bibfnamefont {V.~A.}\ \bibnamefont
  {Mousolou}}\ and\ \bibinfo {author} {\bibfnamefont {E.}~\bibnamefont
  {Sj{\"o}qvist}},\ }\href {https://doi.org/10.1088/1751-8121/aae78b}
  {\bibfield  {journal} {\bibinfo  {journal} {Journal of Physics A:
  Mathematical and Theoretical}\ }\textbf {\bibinfo {volume} {51}},\ \bibinfo
  {pages} {475303} (\bibinfo {year} {2018})}\BibitemShut {NoStop}%
\bibitem [{\citenamefont {Yang}\ \emph {et~al.}(2010)\citenamefont {Yang},
  \citenamefont {Xu}, \citenamefont {Feng},\ and\ \citenamefont
  {Du}}]{Yang2010_NJP12-113039}%
  \BibitemOpen
  \bibfield  {author} {\bibinfo {author} {\bibfnamefont {W.}~\bibnamefont
  {Yang}}, \bibinfo {author} {\bibfnamefont {Z.}~\bibnamefont {Xu}}, \bibinfo
  {author} {\bibfnamefont {M.}~\bibnamefont {Feng}},\ and\ \bibinfo {author}
  {\bibfnamefont {J.}~\bibnamefont {Du}},\ }\href
  {https://doi.org/10.1088/1367-2630/12/11/113039} {\bibfield  {journal}
  {\bibinfo  {journal} {New Journal of Physics}\ }\textbf {\bibinfo {volume}
  {12}},\ \bibinfo {pages} {113039} (\bibinfo {year} {2010})}\BibitemShut
  {NoStop}%
\bibitem [{\citenamefont {Shao}\ \emph {et~al.}(2010)\citenamefont {Shao},
  \citenamefont {Chen}, \citenamefont {Zhang}, \citenamefont {Zhao},\ and\
  \citenamefont {Yeon}}]{Shao2010_EPL90-50003}%
  \BibitemOpen
  \bibfield  {author} {\bibinfo {author} {\bibfnamefont {X.-Q.}\ \bibnamefont
  {Shao}}, \bibinfo {author} {\bibfnamefont {L.}~\bibnamefont {Chen}}, \bibinfo
  {author} {\bibfnamefont {S.}~\bibnamefont {Zhang}}, \bibinfo {author}
  {\bibfnamefont {Y.-F.}\ \bibnamefont {Zhao}},\ and\ \bibinfo {author}
  {\bibfnamefont {K.-H.}\ \bibnamefont {Yeon}},\ }\href
  {https://doi.org/10.1209/0295-5075/90/50003} {\bibfield  {journal} {\bibinfo
  {journal} {{EPL} (Europhysics Letters)}\ }\textbf {\bibinfo {volume} {90}},\
  \bibinfo {pages} {50003} (\bibinfo {year} {2010})}\BibitemShut {NoStop}%
\bibitem [{\citenamefont {Chen}\ \emph {et~al.}(2014)\citenamefont {Chen},
  \citenamefont {Xia},\ and\ \citenamefont {Song}}]{Chen2014_QIP13-1857}%
  \BibitemOpen
  \bibfield  {author} {\bibinfo {author} {\bibfnamefont {Y.-H.}\ \bibnamefont
  {Chen}}, \bibinfo {author} {\bibfnamefont {Y.}~\bibnamefont {Xia}},\ and\
  \bibinfo {author} {\bibfnamefont {J.}~\bibnamefont {Song}},\ }\href
  {https://doi.org/10.1007/s11128-014-0772-4} {\bibfield  {journal} {\bibinfo
  {journal} {Quantum Information Processing}\ }\textbf {\bibinfo {volume}
  {13}},\ \bibinfo {pages} {1857} (\bibinfo {year} {2014})}\BibitemShut
  {NoStop}%
\bibitem [{\citenamefont {Wu}\ \emph {et~al.}(2017)\citenamefont {Wu},
  \citenamefont {Ji},\ and\ \citenamefont {Zhang}}]{Wu2017_SciRep7-46255}%
  \BibitemOpen
  \bibfield  {author} {\bibinfo {author} {\bibfnamefont {J.-L.}\ \bibnamefont
  {Wu}}, \bibinfo {author} {\bibfnamefont {X.}~\bibnamefont {Ji}},\ and\
  \bibinfo {author} {\bibfnamefont {S.}~\bibnamefont {Zhang}},\ }\href
  {https://doi.org/10.1038/srep46255} {\bibfield  {journal} {\bibinfo
  {journal} {Scientific Reports}\ }\textbf {\bibinfo {volume} {7}},\ \bibinfo
  {pages} {46255} (\bibinfo {year} {2017})}\BibitemShut {NoStop}%
\bibitem [{end({\natexlab{b}})}]{endnote:2}%
  \BibitemOpen
  \href@noop {} {} ({\natexlab{b}}),\ \bibinfo {note} {commonly called the
  chiral symmetry in condensed matter physics and high energy physics, the
  anti-symmetry is not a ``symmetry'' in the usual sense because the
  Hamiltonian does not commute with the symmetry operator.}\BibitemShut {Stop}%
\bibitem [{Sup()}]{Supplemental}%
  \BibitemOpen
  \href@noop {} {}\bibinfo {note} {See Supplemental Material at
  \url{http://link.aps.org/supplemental/...}, which includes
  Refs.~[...].}\BibitemShut {Stop}%
\bibitem [{\citenamefont {Purcell}(1946)}]{Purcell46b}%
  \BibitemOpen
  \bibfield  {author} {\bibinfo {author} {\bibfnamefont {E.~M.}\ \bibnamefont
  {Purcell}},\ }\href {https://doi.org/10.1103/physrev.69.674} {\bibfield
  {journal} {\bibinfo  {journal} {Physical Review}\ }\textbf {\bibinfo {volume}
  {69}},\ \bibinfo {pages} {681} (\bibinfo {year} {1946})},\ \bibinfo {note}
  {part of the Proceedings of the American Physical Society}\BibitemShut
  {NoStop}%
\bibitem [{\citenamefont {Berry}(1984)}]{Berry84a}%
  \BibitemOpen
  \bibfield  {author} {\bibinfo {author} {\bibfnamefont {M.~V.}\ \bibnamefont
  {Berry}},\ }\href@noop {} {\bibfield  {journal} {\bibinfo  {journal} {Proc.
  R. Soc. London A}\ }\textbf {\bibinfo {volume} {392}},\ \bibinfo {pages} {45}
  (\bibinfo {year} {1984})}\BibitemShut {NoStop}%
\bibitem [{\citenamefont {Wilczek}\ and\ \citenamefont
  {Zee}(1984)}]{Wilczek1984_PRL52-2111}%
  \BibitemOpen
  \bibfield  {author} {\bibinfo {author} {\bibfnamefont {F.}~\bibnamefont
  {Wilczek}}\ and\ \bibinfo {author} {\bibfnamefont {A.}~\bibnamefont {Zee}},\
  }\href {https://doi.org/10.1103/PhysRevLett.52.2111} {\bibfield  {journal}
  {\bibinfo  {journal} {Phys. Rev. Lett.}\ }\textbf {\bibinfo {volume} {52}},\
  \bibinfo {pages} {2111} (\bibinfo {year} {1984})}\BibitemShut {NoStop}%
\bibitem [{\citenamefont {T{\'o}th}(2007)}]{Toth07a}%
  \BibitemOpen
  \bibfield  {author} {\bibinfo {author} {\bibfnamefont {G.}~\bibnamefont
  {T{\'o}th}},\ }\href {https://doi.org/10.1364/josab.24.000275} {\bibfield
  {journal} {\bibinfo  {journal} {Journal of the Optical Society of America B}\
  }\textbf {\bibinfo {volume} {24}},\ \bibinfo {pages} {275} (\bibinfo {year}
  {2007})}\BibitemShut {NoStop}%
\bibitem [{\citenamefont {T{\'o}th}\ \emph {et~al.}(2009)\citenamefont
  {T{\'o}th}, \citenamefont {Wieczorek}, \citenamefont {Krischek},
  \citenamefont {Kiesel}, \citenamefont {Michelberger},\ and\ \citenamefont
  {Weinfurter}}]{Toth09a}%
  \BibitemOpen
  \bibfield  {author} {\bibinfo {author} {\bibfnamefont {G.}~\bibnamefont
  {T{\'o}th}}, \bibinfo {author} {\bibfnamefont {W.}~\bibnamefont {Wieczorek}},
  \bibinfo {author} {\bibfnamefont {R.}~\bibnamefont {Krischek}}, \bibinfo
  {author} {\bibfnamefont {N.}~\bibnamefont {Kiesel}}, \bibinfo {author}
  {\bibfnamefont {P.}~\bibnamefont {Michelberger}},\ and\ \bibinfo {author}
  {\bibfnamefont {H.}~\bibnamefont {Weinfurter}},\ }\href
  {https://doi.org/10.1088/1367-2630/11/8/083002} {\bibfield  {journal}
  {\bibinfo  {journal} {New Journal of Physics}\ }\textbf {\bibinfo {volume}
  {11}},\ \bibinfo {pages} {083002} (\bibinfo {year} {2009})}\BibitemShut
  {NoStop}%
\bibitem [{\citenamefont {Apellaniz}\ \emph {et~al.}(2015)\citenamefont
  {Apellaniz}, \citenamefont {L{\"u}cke}, \citenamefont {Peise}, \citenamefont
  {Klempt},\ and\ \citenamefont {T{\'o}th}}]{Apellaniz15a}%
  \BibitemOpen
  \bibfield  {author} {\bibinfo {author} {\bibfnamefont {I.}~\bibnamefont
  {Apellaniz}}, \bibinfo {author} {\bibfnamefont {B.}~\bibnamefont
  {L{\"u}cke}}, \bibinfo {author} {\bibfnamefont {J.}~\bibnamefont {Peise}},
  \bibinfo {author} {\bibfnamefont {C.}~\bibnamefont {Klempt}},\ and\ \bibinfo
  {author} {\bibfnamefont {G.}~\bibnamefont {T{\'o}th}},\ }\href
  {https://doi.org/10.1088/1367-2630/17/8/083027} {\bibfield  {journal}
  {\bibinfo  {journal} {New Journal of Physics}\ }\textbf {\bibinfo {volume}
  {17}},\ \bibinfo {pages} {083027} (\bibinfo {year} {2015})}\BibitemShut
  {NoStop}%
\bibitem [{\citenamefont {Pezze}\ \emph {et~al.}(2018)\citenamefont {Pezze},
  \citenamefont {Smerzi}, \citenamefont {Oberthaler}, \citenamefont {Schmied},\
  and\ \citenamefont {Treutlein}}]{Pezze2018_RMP90-035005}%
  \BibitemOpen
  \bibfield  {author} {\bibinfo {author} {\bibfnamefont {L.}~\bibnamefont
  {Pezze}}, \bibinfo {author} {\bibfnamefont {A.}~\bibnamefont {Smerzi}},
  \bibinfo {author} {\bibfnamefont {M.~K.}\ \bibnamefont {Oberthaler}},
  \bibinfo {author} {\bibfnamefont {R.}~\bibnamefont {Schmied}},\ and\ \bibinfo
  {author} {\bibfnamefont {P.}~\bibnamefont {Treutlein}},\ }\href
  {https://doi.org/10.1103/RevModPhys.90.035005} {\bibfield  {journal}
  {\bibinfo  {journal} {Rev. Mod. Phys.}\ }\textbf {\bibinfo {volume} {90}},\
  \bibinfo {pages} {035005} (\bibinfo {year} {2018})}\BibitemShut {NoStop}%
\bibitem [{\citenamefont {Holland}\ and\ \citenamefont
  {Burnett}(1993)}]{Holland93a}%
  \BibitemOpen
  \bibfield  {author} {\bibinfo {author} {\bibfnamefont {M.~J.}\ \bibnamefont
  {Holland}}\ and\ \bibinfo {author} {\bibfnamefont {K.}~\bibnamefont
  {Burnett}},\ }\href {https://doi.org/10.1103/physrevlett.71.1355} {\bibfield
  {journal} {\bibinfo  {journal} {Physical Review Letters}\ }\textbf {\bibinfo
  {volume} {71}},\ \bibinfo {pages} {1355} (\bibinfo {year}
  {1993})}\BibitemShut {NoStop}%
\end{thebibliography}%
% \bibliography{holonomy,physey}

\end{document}